\documentclass{article}
\title{A nonlinear Bloch model for Coulomb interaction in quantum dots}
\author{Brigitte Bidegaray-Fesquet\footnote{Corresponding author. \texttt{Brigitte.Bidegaray@imag.fr}, tel: +33 4 76 51 49 94, fax.: +33 4 76 63 12 63}~ and Kole Keita \\
\small Univ. Grenoble Alpes, LJK, BP 53, 38041 Grenoble Cedex, France\\
\small CNRS, LJK, BP 53, 38041 Grenoble Cedex, France\\
\small \texttt{Brigitte.Bidegaray@imag.fr}, \texttt{Kole.Keita@imag.fr}}
\date{\today}
\usepackage{amsmath,amssymb,amsthm}
\usepackage{float}
\usepackage{graphicx}
\usepackage{tikz}

\def\calC{{\mathcal{C}}}
\def\calE{{\mathcal{E}}}
\def\calM{{\mathcal{M}}}
\def\calN{{\mathcal{N}}}
\def\rmb{{\mathrm{b}}}
\def\rmC{{\mathrm{C}}}
\def\rmc{{\mathrm{c}}}
\def\rmd{{\mathrm{d}}}
\def\rmE{{\mathrm{E}}}
\def\rmF{{\mathrm{F}}}
\def\rmi{{\mathrm{i}}}
\def\rmn{{\mathrm{n}}}
\def\rmv{{\mathrm{v}}}
\def\bfE{{\mathbf{E}}}
\def\bfH{{\mathbf{H}}}
\def\bfL{{\mathbf{L}}}
\def\bfM{{\mathbf{M}}}
\def\bfP{{\mathbf{P}}}
\def\bbC{{\mathbb{C}}}
\def\bbR{{\mathbb{R}}}
\def\a{\alpha}
\def\ap{\alpha'}
\newcommand\ca[1]{c_{#1}}
\newcommand\va[1]{v_{#1}}
\newcommand\cc[1]{c^\dag_{#1}}
\newcommand\vc[1]{v^\dag_{#1}}
\def\Ic{I^\rmc}
\def\Iv{I^\rmv}
\def\Rc{R^\rmc}
\def\Rv{R^\rmv}
\def\Rcv{R^{\rmc-\rmv}}
\def\rhoc{\rho^\rmc}
\def\rhov{\rho^\rmv}
\def\rhocv{\rho^{\rmc\rmv}}
\def\rhovc{\rho^{\rmv\rmc}}
\def\eps{\varepsilon}

\def\HC{{H^\rmC}}

\newcommand\mean[1]{\left\langle#1\right\rangle}
\def\diag{\operatorname{diag}}
\def\Tr{\operatorname{Tr}}
\def\curl{\operatorname{curl}}
\def\div{\operatorname{div}}
\def\card{\operatorname{card}}
\def\sech{\operatorname{sech}}
\def\WT{\stackrel{{\rm (WT)}}{=}}

\newtheorem{Theorem}{Theorem}
\theoremstyle{definition}
\newtheorem{Definition}{Definition}

\newtheorem{Property}{Property}
\theoremstyle{remark}
\newtheorem{Remark}{Remark}

\begin{document}
\maketitle

\begin{abstract}
In this paper, we first derive a Coulomb Hamiltonian for electron--electron interaction in quantum dots in the Heisenberg picture. 
Then we use this Hamiltonian to enhance a Bloch model, which happens to be nonlinear in the density matrix. 
The coupling with Maxwell equations in case of interaction with an electromagnetic field is also considered from the Cauchy problem point of view. The study is completed by numerical results and a discussion about the advisability of neglecting intra-band coherences, as is done in part of the literature. 
\end{abstract}


\section{Introduction}

Bloch model is a very common model to describe the time evolution of a system of electrons in different contexts such as gases of electrons, glasses, or crystals.
The very classical case of gases and glasses involves isotropic media. 
The electrons are supposed to be localized and non interacting. 
Their behavior is averaged at the mesoscopic scale.
This leads to relatively simple models where matter energy levels are quantized and labelled by integers.
The case of crystals \cite{Besse-BidegarayFesquet-Bourgeade-Degond-Saut04}
also involves integer indexed levels, but symmetries and directions  in matter have to be taken into account.

Bloch model has also been extended to the description of quantum wells \cite{Hess-Kuhn96a,Kuhn-Rossi92,Haug-Koch09}, and quantum dots \cite{Gehrig-Hess02,BidegarayFesquet10}. 
In these models, matter is described by the state of two species of particles (electrons and holes, or equivalently conduction and valence electrons). 
In quantum wells, energy levels are indexed by vectors, which correspond to displacements over the underlying lattice.
In contrast, the confinement of electrons in quantum dots leads to integer indexed levels like in gases, which often leads to consider quantum dots as pseudo-atoms, but this is a very raw vision.
In particular, among other differences, electrons at the same mesoscopic location do interact directly \textit{via} Coulomb interaction.

\cite{BidegarayFesquet10} is a preliminary paper that derives basic Bloch equations for two species of electrons (conduction and valence) only taking into account the free electron Hamiltonian and the interaction with a laser electric field.
The aim of the present paper is to include properly Coulomb interaction in this model.
Beyond the sole derivation of the model, we want in particular to study its mathematical properties.
In the continuation of \cite{Bidegaray01b,Bidegaray-Bourgeade-Reignier01}, we want to show that a certain number of properties are preserved through the time evolution, such as Hermicity and positiveness of the density matrix.

\subsection{Outline}

The outline of this paper is as follows. 
We devote Section \ref{sec:basicmodel} to the description of the basic Bloch model which does not include Coulomb interaction but only the free energies of the electrons and the action of an electromagnetic field.
In Section \ref{sec:Coulomb}, we derive the Coulomb Hamiltonian in terms of the conduction and valence operators. 
The associated Heisenberg equation is derived in Section \ref{sec:Heisenberg}, but it ends up with an open system of equations. 
The system is closed using the Wick theorem, and the final Bloch equation has a Liouville form, but is nonlinear.
This nonlinearity does not allow to use previous literature directly and  we prove anew the Hermicity, positiveness and boundedness results in Section \ref{sec:maths}.
In Section \ref{sec:numerics}, we show the impact of the Coulomb contribution in numerical results and also compare our model with a vanishing intra-band coherence model defined in \cite{Gehrig-Hess02}.

\subsection{The basic Bloch model}
\label{sec:basicmodel}

Let us first recall the main results obtained in \cite{BidegarayFesquet10} and fix the notations.

\subsubsection{Commutators and Heisenberg equation}

Let $A$ and $B$ be two operators, we define their commutator by $[A,B]=AB-BA$ and their skew-commutator by $\{A,B\}=AB+BA$.
For an operator $A$, we define the associated observable $\langle A\rangle = \Tr(S_0 A)$ by averaging with respect to the initial state density $S_0$ of the system.
If the system is described by a Hamiltonian $H$, the time-evolution for this observable is given by the Heisenberg equation
\begin{equation}
\label{eq:Heisenberg}
\rmi\hbar \partial_t \langle A \rangle = \langle [A,H] \rangle, 
\end{equation}
where $\hbar$ is the reduced Planck constant.
When the observable is the density matrix, the Heisenberg equation of motion is called the Bloch equation.

\subsubsection{Operators for quantum dots}
\label{sec:operators}

A quantum dot is defined as a collection of conduction and valence electrons.
There is of course no conduction in quantum dots since the electrons are confined in every direction, but this terminology is useful to distinguish between the valence electrons --- which are in fact the absence of holes in the valence band --- and the free, but confined, electrons.
For each species, energy levels are quantized and indexed by a set of integers, $\Ic$ and $\Iv$, for conduction and valence electrons, respectively.
For $i\in\Ic$, we define the creation and annihilation operators $\cc{i}$ and $\ca{i}$. Likewise, for valence electrons, we define the creation and annihilation operators $\vc{i}$ and $\va{i}$.

\begin{Property}
\label{prop:skew}
\begin{equation*}
\{c_i,c_j^\dag\}=\delta_{i,j},
\hspace{1cm}
\{c_i,c_j\}=\{c_i^\dag,c_j^\dag\}=0,
\end{equation*}
\begin{equation*}
\{v_i,v_j^\dag\}=\delta_{i,j},
\hspace{1cm}
\{v_i,v_j\}=\{v_i^\dag,v_j^\dag\}=0,
\end{equation*}
where $\delta_{i,j}$ denotes the Kronecker symbol.
\end{Property}

This implies in particular that $\ca{i} \ca{i} = \cc{i} \cc{i} = 0$, which means that it is impossible to create twice or annihilate twice the same electron. This is the Pauli exclusion rule: electrons are fermions.
Of course any conduction operator commutes with any valence operator.

\subsubsection{Observables for quantum dots}

The observable we are interested in is the \textit{density matrix}.
It includes a \textit{conduction density matrix}, which elements are the $\rhoc_{ij}=\langle\cc{j}\ca{i}\rangle$. 
This matrix is Hermitian and positive semi-definite.
Its diagonal terms $\rhoc_{ii}=\langle \cc{i}\ca{i}\rangle$ are also called \textit{populations} and give the probability to find an electron in state $i$.
The off-diagonal terms, $\rhoc_{ij}$, $i\neq j$ are called \textit{(intra-band) coherences}.
Of course we also define a \textit{valence density matrix}, which elements are the $\rhov_{ij}=\langle\vc{j}\va{i}\rangle$. 
Besides we are interested in \textit{inter-band coherences} defined by $\rhocv_{ij}=\langle \vc{j}\ca{i}\rangle$.
The entries of these matrices are the variables of the Bloch equation. 
They are cast in a single density matrix
\begin{equation*}
\rho=\left(\begin{array}{cc}
\rhoc & \rhocv \\
\rhovc & \rhov 
\end{array}\right)
\end{equation*}
where $\rhovc={\rhocv}^*$, which ensures that $\rho$ is Hermitian and positive semi-definite.

\subsubsection{Free electron Hamiltonians and interaction with a laser}

The basic Bloch equation for quantum dot is derived in \cite{BidegarayFesquet10}.
It takes into account two types of Hamiltonians in the Heisenberg equation for the density matrix, namely free electron Hamiltonians and interaction Hamiltonians with a laser field.
The free electron Hamiltonians read
\begin{equation*}
H^\rmc_0 = \sum_{k\in\Ic} \epsilon^\rmc_k \cc{k}\ca{k},
\hspace{1cm}
H^\rmv_0 = \sum_{k\in\Iv} \epsilon^\rmv_k \vc{k}\va{k},
\end{equation*}
for conduction and valence electrons respectively. The coefficients $\epsilon^\rmc_k$ and $\epsilon^\rmv_k$ are the free electron energies associated with each electron level. The electron levels are described by the wave functions $\psi_k^{\rmc}$ and $\psi_k^{\rmv}$, solutions to the free electron Schr\"odinger equation subject to the boundary conditions of the quantum dot (see \cite{Haug-Koch09}).
The interaction with a laser characterized by its time-dependent electric field $\bfE(t)$ is described by the Hamiltonians
\begin{eqnarray*}
H^{\rm Lc} 
& = & \frac12 \sum_{(k,l)\in (\Ic)^2} ( \bfE(t) \cdot \bfM_{kl}^\rmc \cc{k}\ca{l} + \bfE^*(t) \cdot \bfM_{kl}^{\rmc*} \cc{l}\ca{k} ), \\
H^{\rm Lv} 
& = & \frac12 \sum_{(k,l)\in (\Iv)^2} ( \bfE(t) \cdot \bfM_{kl}^\rmv \vc{k}\va{l} + \bfE^*(t) \cdot \bfM_{kl}^{\rmv*} \vc{l}\va{k} ), \\
H^{\rm Lcv} 
& = & \sum_{(k,l)\in \Ic\times \Iv} ( \bfE(t) \cdot \bfM_{kl}^{\rmc\rmv} \cc{k}\va{l} + \bfE^*(t) \cdot \bfM_{kl}^{\rmc\rmv*} \vc{l}\ca{k} ),
\end{eqnarray*}
where the dipolar moment matrices are matrices with entries in $\bbC^3$ and may be expressed in terms of the wave functions associated to each level: 
\begin{eqnarray*}
\bfM_{kl}^\rmc & = & \int \rmd r\ \psi_l^{\rmc*}(r)\,er\,\psi_k^{\rmc}(r), \\
\bfM_{kl}^\rmv & = & \int \rmd r\ \psi_l^{\rmv*}(r)\,er\,\psi_k^{\rmv}(r), \\
\bfM_{kl}^{\rmc\rmv} & = & \int \rmd r\ \psi_l^{\rmv*}(r)\,er\,\psi_k^{\rmc}(r),
\end{eqnarray*}
where $e$ is the unsigned charge of the electron.
Injecting these Hamiltonians in the Heisenberg equation, the basic Bloch equations can be cast in Liouville form
\begin{equation}
\label{eq:basicBloch}
\rmi\hbar\partial_t \rho = [V_0(t),\rho].
\end{equation}
In equation \eqref{eq:basicBloch}, $V_0(t)=V^\rmF+V^\rmE(\bfE(t))$ is a sum of a constant term $V^\rmF$ stemming from the free energies collected in diagonal matrices $E_0^\rmc=\diag(\{\epsilon_i^\rmc\}_{i\in\Ic})$ and $E_0^\rmv=\diag(\{\epsilon_i^\rmv\}_{i\in\Iv})$, and a time dependent term due to the interaction with the electric field:
\begin{equation*}
V^\rmF = \left(\begin{array}{cc}
E_0^\rmc & 0 \\
0 & E_0^\rmv \\
\end{array}\right)
\text{ and }
V^\rmE(\bfE(t)) = \left(\begin{array}{cc}
\Re\bfE(t)\cdot\bfM^\rmc & \bfE(t)\cdot\bfM^{\rmc\rmv} \\
\bfE^*(t)\cdot\bfM^{\rmc\rmv*} & \Re\bfE(t)\cdot\bfM^\rmv \\
\end{array}\right).
\end{equation*}
The scalar product of the electrical field and the dipolar moment matrix, is a scalar product in $\bbC^3$ which yields a matrix with entries in $\bbC$, with the same dimension as $\rho$, as is necessary to give a meaning to the right-hand side of \eqref{eq:basicBloch}.

\subsubsection{Mathematical properties of the Liouville equation}

Equation \eqref{eq:basicBloch} clearly preserves the Hermitian structure of $\rho$.
Its exact solution is 
\begin{equation}
\label{eq:exact}
\rho(t) = \exp\left(-\frac{\rmi}{\hbar} \int_0^t V_0(\tau)\ \rmd\tau\right) \rho(0) \exp\left(\frac{\rmi}{\hbar} \int_0^t V_0(\tau)\ \rmd\tau\right).
\end{equation}
This expression allows to prove a certain number of properties (see \cite{BidegarayFesquet10}). Let $d=\card(\Ic)+\card(\Iv)$ be the total number of levels, given a positive semi-definite initial data $\rho(0)\in\calM_d(\bbC)$ and a continuous electric field $\bfE(t)$, 
\begin{itemize}
\item equation \eqref{eq:basicBloch} is globally well-posed, i.e. there exists a unique solution $\rho\in\calC^1(\bbR^+;\calM_d(\bbC))$ that exists for all times $t\geq0$ and which depends continuously on the data (parameters, initial data); 
\item for all time $\rho(t)$ is a positive semi-definite matrix;
\item its trace is conserved through the time evolution.
\end{itemize}

\section{Second quantification Coulomb Hamiltonian}
\label{sec:Coulomb}

Coulomb interaction can  be introduced using field operators.
We denote by $\hat\psi_\rmc^\dag(r)$ and $\hat\psi_\rmv^\dag(r)$ the creation field-operators of respectively a conduction electron and a valence electron at the space location $r$, and $\hat\psi_\rmc(r)$ and $\hat\psi_\rmv(r)$ the corresponding annihilation field-operators.
We consider that there are $N$ relevant electrons in the quantum dot, and can write the Coulomb Hamiltonians as 
\begin{subequations}
\label{eq:Coulomb_initial}
\begin{equation}
H^{\rmc-\rmc} = \frac12 \sum_{i,j=1}^N \iint \rmd r_i \rmd r_j\ \hat\psi_\rmc^\dag(r_i) \hat\psi_\rmc^\dag(r_j) V^\rmc(r_i,r_j) \hat\psi_\rmc(r_j) \hat\psi_\rmc(r_i),
\end{equation}
\begin{equation}
H^{\rmv-\rmv} = \frac12 \sum_{i,j=1}^N \iint \rmd r_i \rmd r_j\ \hat\psi_\rmv(r_i) \hat\psi_\rmv(r_j) V^\rmv(r_i,r_j) \hat\psi_\rmv^\dag(r_j) \hat\psi_\rmv^\dag(r_i),
\end{equation}
\begin{equation}
H^{\rmc-\rmv} = \sum_{i,j=1}^N \iint \rmd r_i \rmd r_j\ \hat\psi_\rmc^\dag(r_i) \hat\psi_\rmv(r_j) V^{\rmc-\rmv}(r_i,r_j) \hat\psi_\rmv^\dag(r_j) \hat\psi_\rmc(r_i),
\end{equation}
\end{subequations}
where $V^\rmc$, $V^\rmv$ and $V^{\rmc-\rmv}$ are the conduction--conduction, valence--valence and conduction--valence Coulomb potentials.
The Coulomb potentials have the form 
\begin{equation*}
V^\rmc(r,r') = V^\rmv(r,r') = \frac{k_C}{|r-r'|} 
\text{ and }
V^{\rmc-\rmv}(r,r') = -\frac{k_C}{|r-r'|},
\end{equation*}
where $k_C$ is Coulomb's constant. 
The difference of treatment of conduction and valence electrons stems from the fact that Coulomb interaction describes the interaction of electrons and holes (see e.g. \cite{Haug-Koch09}) and that the presence of an electron in the valence band is indeed the absence of the corresponding hole, and vice-versa, which inverts the role of creation and annihilation operators. 
The total Coulomb Hamiltonian is $\HC=H^{\rmc-\rmc}+H^{\rmv-\rmv}+H^{\rmc-\rmv}$.

We want to derive Bloch-type equations including the Coulomb interaction. 
Bloch equations have the advantage not to depend explicitly on the exact form of the field-operators. 
To this aim, we write
\begin{subequations}
\label{eq:field_ops}
\begin{equation}
\hat\psi_\rmc(r) = \sum_{\a\in\Ic} \psi_\a^\rmc(r) \ca{\a},
\hspace{1cm}
\hat\psi_\rmc^\dag(r) = \sum_{\a\in\Ic} \psi_\a^{\rmc*}(r) \cc{\a},
\end{equation}
\begin{equation}
\hat\psi_\rmv(r) = \sum_{\a\in\Iv} \psi_\a^\rmv(r) \va{\a},
\hspace{1cm}
\hat\psi_\rmv^\dag(r) = \sum_{\a\in\Iv} \psi_\a^{\rmv*}(r) \vc{\a}.
\end{equation}
\end{subequations}
They are weighted by the conduction and valence electron wave functions $\psi_\a^\rmc(r)$ and $\psi_\a^\rmv(r)$, which are the same as those who occurred in the expression of the dipolar moment matrices.
In the sequel, to avoid unnecessary written complexity, we will often omit to specify which set the indices belong to.

Inserting decompositions \eqref{eq:field_ops} in Hamiltonians \eqref{eq:Coulomb_initial} we obtain
\begin{subequations}
\label{eq:Coulomb_eh}
\begin{equation}
H^{\rmc-\rmc} = \sum_{\substack{\a_1,\a_2,\\ \ap_1,\ap_2}} R_{\a_1\a_2\ap_1\ap_2}^\rmc \cc{\a_1} \cc{\a_2} \ca{\ap_2} \ca{\ap_1},
\end{equation}
\begin{equation}
H^{\rmv-\rmv} = \sum_{\substack{\a_1,\a_2,\\ \ap_1,\ap_2}} R_{\a_1\a_2\ap_1\ap_2}^\rmv \va{\ap_1} \va{\ap_2} \vc{\a_2} \vc{\a_1},
\end{equation}
\begin{equation}
H^{\rmc-\rmv} = - \sum_{\substack{\a_1,\a_2,\\ \ap_1,\ap_2}} R_{\a_1\a_2\ap_1\ap_2}^{\rmc-\rmv} \cc{\a_1} \va{\ap_2} \vc{\a_2} \ca{\ap_1},
\end{equation}
\end{subequations}
where
\begin{subequations}
\label{eq:R}
\begin{equation}
R_{\a_1\a_2\ap_1\ap_2}^\rmc = \frac{N^2}2 \iint \rmd r \rmd r'\ \psi_{\a_1}^{\rmc*}(r) \psi_{\a_2}^{\rmc*}(r') V^\rmc(r,r') \psi_{\ap_2}^\rmc(r') \psi_{\ap_1}^\rmc(r),
\end{equation}
\begin{equation}
R_{\ap_1\ap_2\a_1\a_2}^\rmv = \frac{N^2}2 \iint \rmd r \rmd r'\ \psi_{\ap_1}^\rmv(r) \psi_{\ap_2}^\rmv(r') V^\rmv(r,r') \psi_{\a_2}^{\rmv*}(r') \psi_{\a_1}^{\rmv*}(r),
\end{equation}
\begin{equation}
R_{\a_1\a_2\ap_1\ap_2}^{\rmc-\rmv} = - N^2 \iint \rmd r \rmd r'\ \psi_{\a_1}^{\rmc*}(r) \psi_{\ap_2}^\rmv(r') V^{\rmc-\rmv}(r,r') \psi_{\a_2}^{\rmv*}(r') \psi_{\ap_1}^\rmc(r).
\end{equation}
\end{subequations}
The symmetries in the integrands of \eqref{eq:R} induce the following properties.

\begin{Property}
\label{prop:R}
Since $V(r,r')$ is an even function of $r-r'$, variables $r$ and $r'$ play the same role and
\begin{equation*}
R_{\a_1\a_2\ap_1\ap_2}^\rmc = R_{\a_2\a_1\ap_2\ap_1}^\rmc,
\hspace{1cm}
R_{\a_1\a_2\ap_1\ap_2}^\rmv = R_{\a_2\a_1\ap_2\ap_1}^\rmv.
\end{equation*}
Since $V(r,r')$ is a real valued function
\begin{equation*}
R_{\a_1\a_2\ap_1\ap_2}^\rmc = R_{\ap_1\ap_2\a_1\a_2}^{\rmc*},
\hspace{1cm}
R_{\a_1\a_2\ap_1\ap_2}^\rmv = R_{\ap_1\ap_2\a_1\a_2}^{\rmv*},
\end{equation*}
\begin{equation*}
R_{\a_1\a_2\ap_1\ap_2}^{\rmc-\rmv} = \left(R_{\ap_1\ap_2\a_1\a_2}^{\rmc-\rmv}\right)^*.
\end{equation*}
\end{Property}

The Pauli exclusion principle (skew-commutation, Property \ref{prop:skew}) also induces that some terms in $H^{\rmc-\rmc}$ and $H^{\rmv-\rmv}$ are necessarily zero.

\begin{Property}
\label{prop:exclusion}
If $\a_1=\a_2$ or $\ap_1=\ap_2$,
\begin{equation*}
\cc{\a_1} \cc{\a_2} \ca{\ap_2} \ca{\ap_1} = 0,
\hspace{1cm}
\va{\ap_1} \va{\ap_2} \vc{\a_2} \vc{\a_1} = 0.
\end{equation*}
\end{Property}

The order of the operators (which has to be read from the right to the left) in the Coulomb Hamiltonians \eqref{eq:Coulomb_initial} has a meaning: in order that two particles interact via Coulomb interaction they have to pre-exist at locations $r_i$ and $r_j$. 
Then they are annihilated while interacting and recreated at the same locations.

\begin{Definition}
\label{def:normalorder}
A product of operators will be said to be in the \textit{normal order}, if the annihilation operators are on the right and the creation operators on the left.
\end{Definition}

$H^{\rmc-\rmc}$ already follows a normal ordered form and we can keep it untouched:
\begin{subequations}
\label{eq:Coulomb_cv}
\begin{equation}
H^{\rmc-\rmc} = \sum_{\substack{\a_1,\a_2,\\ \ap_1,\ap_2}} R_{\a_1\a_2\ap_1\ap_2}^\rmc \cc{\a_1} \cc{\a_2} \ca{\ap_2} \ca{\ap_1}.
\end{equation}

Hamiltonians $H^{\rmv-\rmv}$ and $H^{\rmc-\rmv}$ given by equations (\ref{eq:Coulomb_eh}b) and (\ref{eq:Coulomb_eh}c) do not follow normal ordered forms.

To express $H^{\rmv-\rmv}$ we need to compute a normal ordered form of $\va{\ap_1} \va{\ap_2} \vc{\a_2} \vc{\a_1}$: 
\begin{eqnarray*}
\va{\ap_1} \va{\ap_2} \vc{\a_2} \vc{\a_1}
& = & \delta_{\a_1\ap_1} \delta_{\a_2\ap_2} - \delta_{\a_1\ap_1} \vc{\a_2} \va{\ap_2}
- \delta_{\ap_2\a_1} \delta_{\ap_1\a_2}
+ \delta_{\ap_2\a_1} \vc{\a_2} \va{\ap_1} \\
&& + \delta_{\ap_1\a_2} \vc{\a_1} \va{\ap_2} 
- \delta_{\a_2\ap_2} \vc{\a_1} \va{\ap_1}
+ \vc{\a_1} \vc{\a_2} \va{\ap_2} \va{\ap_1}.
\end{eqnarray*}

Thanks to Property \ref{prop:R}, $R_{\a_1\a_2\ap_1\ap_2}^\rmv = R_{\a_2\a_1\ap_2\ap_1}^\rmv$, therefore $- \delta_{\a_1\ap_1} \vc{\a_2} \va{\ap_2}$ and $- \delta_{\a_2\ap_2} \vc{\a_1} \va{\ap_1}$ lead to the same contribution.
The same argument can be applied to $\delta_{\ap_2\a_1} \vc{\a_2} \va{\ap_1}$ and $\delta_{\ap_1\a_2} \vc{\a_1} \va{\ap_2}$.
Hence
\begin{equation}
H^{\rmv-\rmv} = 
2 \sum_{\a,\ap,\beta} (R_{\beta\a\ap\beta}^\rmv - R_{\beta\a\beta\ap}^\rmv) \vc{\a} \va{\ap}
+ \sum_{\substack{\a_1,\a_2,\\ \ap_1,\ap_2}} R_{\a_1\a_2\ap_1\ap_2}^\rmv 
\vc{\ap_1} \vc{\ap_2} \va{\a_2} \va{\a_1}.
\end{equation}
In the definition of $H^{\rmv-\rmv} $ we have dropped the $\delta_{\a_1\ap_1} \delta_{\a_2\ap_2}$ and $- \delta_{\ap_2\a_1} \delta_{\ap_1\a_2}$ terms which would lead to zero contributions in the Heisenberg equation.

In the same way
$\cc{\a_1} \va{\ap_2} \vc{\a_2} \ca{\ap_1}
= \delta_{\a_2\ap_2} \cc{\a_1} \ca{\ap_1} 
- \cc{\a_1} \vc{\a_2} \va{\ap_2} \ca{\ap_1}$,
hence
\begin{equation}
H^{\rmc-\rmv} = - \sum_{\a,\ap,\beta}  R_{\a\beta\ap\beta}^{\rmc-\rmv} 
\cc{\a} \ca{\ap}
+ \sum_{\substack{\a_1,\a_2,\\ \ap_1,\ap_2}} R_{\a_1\a_2\ap_1\ap_2}^{\rmc-\rmv} \cc{\a_1} \vc{\a_2} \va{\ap_2} \ca{\ap_1}.
\end{equation}
\end{subequations}

\section{Formulation of the Heisenberg equation}
\label{sec:Heisenberg}

We now write Heisenberg equation \eqref{eq:Heisenberg} where $A$ are operators $\cc{j} \ca{i}$, $\vc{j} \va{i}$ or $\vc{j} \ca{i}$ and $H$ are the Hamiltonians defined by Equation \eqref{eq:Coulomb_cv}.

\subsection{Wick theorem}

For the free Hamiltonians and the laser interactions, the computation of the commutators led to expressions in terms of the two-operator densities, and therefore to a closed set of equations \cite{BidegarayFesquet10}.
This will not be the case any more here, since the commutators stemming from Coulomb Hamiltonians will give rise to four-operator densities. 
To go further we should \textit{a priori} have evolution equations for these observables \textit{via} the Heisenberg equation, computing commutators with the already defined Coulomb Hamiltonians. 
This would lead inevitably to six-operator densities, and so on. 
To avoid this endless procedure, we have to close the system at some point. 
This is the goal of the Wick theorem \cite{Wick50} which amounts in our case to write the four-operator densities as sums of products of two-operator densities following e.g. the rule
\begin{equation*}
\mean{\cc{\a_1} \cc{\a_2} \ca{\ap_2} \ca{\ap_1}} 
\WT \mean{\cc{\a_1} \ca{\ap_1}} \mean{\cc{\a_2} \ca{\ap_2}} - \mean{\cc{\a_1} \ca{\ap_2}} \mean{\cc{\a_2} \ca{\ap_1}},
\end{equation*}
where the symbol $\WT$ means "is approximated through Wick theorem".

\subsection{Commutators involving $\cc{j} \ca{i}$}
\label{sec:c}

In order to derive the Heisenberg equation, we have to compute $[\cc{j} \ca{i},H^{\rmc-\rmc}]$ and $[\cc{j} \ca{i},H^{\rmc-\rmv}]$ (since $[\cc{j} \ca{i},H^{\rmv-\rmv}]$ is clearly zero).

\subsubsection{Commutator with $H^{\rmc-\rmc}$}
\label{sec:details}

According to Equation (\ref{eq:Coulomb_cv}a)  
\begin{equation*}
[\cc{j} \ca{i},H^{\rmc-\rmc}] 
= \sum_{\substack{\a_1,\a_2,\\ \ap_1,\ap_2}} 
\Rc_{\a_1\a_2\ap_1\ap_2} [\cc{j} \ca{i},\cc{\a_1} \cc{\a_2} \ca{\ap_2} \ca{\ap_1}].
\end{equation*}

\begin{Remark}
We already know many situations where $[\cc{j} \ca{i},\cc{\a_1} \cc{\a_2} \ca{\ap_2} \ca{\ap_1}]$ is necessarily zero:
\begin{itemize}
\item if none of the indices $\a_1,\a_2,\ap_1,\ap_2$ is equal either to $i$ or $j$,
\item if $\a_1=\a_2$ or $\ap_1=\ap_2$ (see Property \ref{prop:exclusion}).
\end{itemize}
\end{Remark}

We compute separately each commutator:
\begin{eqnarray*}
[\cc{j} \ca{i},\cc{\a_1} \cc{\a_2} \ca{\ap_2} \ca{\ap_1}] 
& = & \delta_{i\a_1} \cc{j} \cc{\a_2} \ca{\ap_2} \ca{\ap_1} 
+ \delta_{i\a_2} \cc{j} \cc{\a_1} \ca{\ap_1} \ca{\ap_2} \\
&& - \delta_{j\ap_2} \cc{\a_2} \cc{\a_1} \ca{\ap_1} \ca{i}
- \delta_{j\ap_1} \cc{\a_1} \cc{\a_2} \ca{\ap_2} \ca{i}.
\end{eqnarray*}
Using Property \ref{prop:R}, we see that
\begin{eqnarray*}
[\cc{j} \ca{i},H^{\rmc-\rmc}] 
& = & \sum_{\substack{\a_1,\a_2,\\ \ap_1,\ap_2}} \Rc_{\a_1\a_2\ap_1\ap_2} \left(2 \delta_{i\a_1} \cc{j} \cc{\a_2} \ca{\ap_2} \ca{\ap_1} 
- 2 \delta_{j\ap_1} \cc{\a_1} \cc{\a_2} \ca{\ap_2} \ca{i}\right)\\
& = & 2 \sum_{\substack{\a,\\ \ap_1,\ap_2}} 
\Rc_{i\a\ap_1\ap_2} \cc{j} \cc{\a} \ca{\ap_2} \ca{\ap_1} 
- 2 \sum_{\substack{\a_1,\a_2,\\ \ap}} 
\Rc_{\a_1\a_2j\ap} \cc{\a_1} \cc{\a_2} \ca{\ap} \ca{i}.
\end{eqnarray*}
We now apply Wick theorem which leads to
\begin{eqnarray*}
\mean{[\cc{j} \ca{i},H^{\rmc-\rmc}]} 
& \WT & 2 \sum_{\substack{\a,\\ \ap_1,\ap_2}} 
\Rc_{i\a\ap_1\ap_2} \left(\mean{\cc{j}\ca{\ap_1}}\mean{\cc{\a}\ca{\ap_2}}- \mean{\cc{j}\ca{\ap_2}}\mean{\cc{\a}\ca{\ap_1}} \right) \\
&& - 2 \sum_{\substack{\a_1,\a_2,\\ \ap}} 
\Rc_{\a_1\a_2j\ap} \left(\mean{\cc{\a_1}\ca{i}}\mean{\cc{\a_2}\ca{\ap}}-\mean{\cc{\a_1}\ca{\ap}}\mean{\cc{\a_2}\ca{i}}\right) \\
& = & 2 \sum_{\substack{k, \\ \a,\ap}} 
\left(\Rc_{i\a k\ap}-\Rc_{i\a\ap k}\right) \mean{\cc{\a}\ca{\ap}} \mean{\cc{j}\ca{k}}\\
&& - 2 \sum_{\substack{k,\\ \a,\ap}} 
\left(\Rc_{k\a j\ap}-\Rc_{\a kj\ap}\right) \mean{\cc{\a}\ca{\ap}} \mean{\cc{k}\ca{i}}.
\end{eqnarray*}
Defining matrix $\Lambda^\rmc(\rho)$ as
\begin{equation}
\label{eq:Lambdac}
\Lambda_{ik}^\rmc(\rho) = 2 \sum_{(\a,\ap)\in(\Ic)^2} 
\left(\Rc_{i\a k\ap}-\Rc_{i\a\ap k}\right) \rhoc_{\ap\a},
\end{equation}
we can cast the result as
\begin{equation*}
\mean{[\cc{j} \ca{i},H^{\rmc-\rmc}]} \WT [\Lambda^\rmc(\rho),\rhoc]_{ij}.
\end{equation*}

\subsubsection{Commutator with $H^{\rmc-\rmv}$}

The same sort of computation as in Section \ref{sec:details} is performed on  Equation (\ref{eq:Coulomb_cv}c) to compute $[\cc{j} \ca{i},H^{\rmc-\rmv}]$.
 
Using the matrices $\zeta^\rmv(\rho)$, $\gamma^{\rmc-\rmv}(\rho)$, and $\eta_{ik}^{\rmc-\rmv}$ where
\begin{eqnarray}
\label{eq:zetav}
\zeta_{ik}^\rmv(\rho) & = & \sum_{(\a,\ap)\in(\Iv)^2} \Rcv_{i\a k\ap}\rhov_{\ap\a}, \\
\label{eq:gammacv}
\gamma_{ik}^{\rmc-\rmv}(\rho) & = & -\sum_{(\a,\ap)\in\Iv\times\Ic} \Rcv_{i\a\ap k}\rhocv_{\ap\a}, \\
\label{eq:etacv}
\eta_{ik}^{\rmc-\rmv} & = & - \sum_{\beta\in\Iv} \Rcv_{i\beta k\beta},
\end{eqnarray}
we obtain that 
\begin{equation*}
\mean{[\cc{j} \ca{i},H^{\rmc-\rmv}]}
\WT [\zeta^\rmv(\rho)+\eta^{\rmc-\rmv},\rhoc]_{ij} + \sum_k \gamma_{ik}^{\rmc-\rmv}(\rho) \rhovc_{kj} 
- \sum_k \rhocv_{ik} \gamma_{kj}^{\rmc-\rmv*}(\rho).
\end{equation*}

\subsection{Commutators involving $\vc{j} \va{i}$}
\label{sec:v}

\subsubsection{Commutator with $H^{\rmv-\rmv}$}

According to Equation (\ref{eq:Coulomb_cv}b), we have to evaluate two types of commutators to compute $[\vc{j} \va{i},H^{\rmv-\rmv}]$.
The first commutator is clearly computed in the same way as $[\cc{j} \ca{i},H^{\rmc-\rmc}]$ replacing conduction electron operators by valence ones.
For the second part, we have to compute $[\vc{j} \va{i},\vc{\a} \va{\ap}]$.
We obtain
\begin{equation*}
\mean{[\vc{j} \va{i},H^{\rmv-\rmv}]} 
\WT [\Lambda^\rmv(\rho)+\kappa^\rmv,\rhov]_{ij},
\end{equation*}
where $\Lambda^\rmv(\rho)$ and $\kappa^\rmv$ are defined as
\begin{eqnarray}
\label{eq:Lambdav}
\Lambda_{ik}^\rmv(\rho) & = & 2 \sum_{(\a,\ap)\in(\Iv)^2} 
\left(\Rv_{i\a k\ap}-\Rv_{i\a\ap k}\right) \rhov_{\ap\a}, \\
\label{eq:kappav}
\kappa_{ik}^\rmv & = & 2 \sum_{\beta\in\Iv} (\Rv_{\beta ik\beta}-\Rv_{\beta i\beta k}).
\end{eqnarray}

\subsubsection{Commutator with $H^{\rmc-\rmv}$}

According to Equation (\ref{eq:Coulomb_cv}c), the commutator $[\vc{j} \va{i},H^{\rmc-\rmv}]$ only involves the first term of $H^{\rmc-\rmv}$, which we can write as
\begin{equation*}
[\vc{j} \va{i},H^{\rmc-\rmv}] 
= \sum_{\substack{\a_1,\a_2,\\ \ap_1,\ap_2}} 
\Rcv_{\a_1\a_2\ap_1\ap_2} \cc{\a_1}  \ca{\ap_1} [\vc{j} \va{i},\vc{\a_2} \va{\ap_2}].
\end{equation*}
The simplifaction of the commutators and the Wick theorem leads to
\begin{equation*}
\mean{[\vc{j} \va{i},H^{\rmc-\rmv}]}
\WT [\zeta^\rmc(\rho),\rhov]_{ij} + \sum_k \gamma_{ik}^{\rmc-\rmv*}(\rho) \rhocv_{kj} - \sum_k \rhovc_{ik}\gamma_{kj}^{\rmc-\rmv}(\rho),
\end{equation*}
using the previously defined $\gamma^{\rmc-\rmv}(\rho)$ and the new matrix $\zeta^\rmc(\rho)$ defined by
\begin{equation}
\label{eq:zetac}
\zeta_{ik}^\rmc(\rho) = \sum_{(\a,\ap)\in(\Ic)^2} \Rcv_{\a i\ap k}\rhoc_{\ap\a}.
\end{equation}

\subsection{Commutators involving $\vc{j} \ca{i}$}
\label{sec:cv}

The commutators involving $\vc{j} \ca{i}$ are all possible to write with already defined matrices, indeed we first compute $[\vc{j} \ca{i},H^{\rmc-\rmc}]$ according to Equation (\ref{eq:Coulomb_cv}a) and recognize 
\begin{equation*}
\mean{[\vc{j} \ca{i},H^{\rmc-\rmc}]} 
\WT \sum_k \Lambda_{ik}^\rmc(\rho) \rhocv_{kj}.
\end{equation*}
Then Equation (\ref{eq:Coulomb_cv}b) leads to 
\begin{equation*}
\mean{[\vc{j} \ca{i},H^{\rmv-\rmv}]} 
\WT - \sum_k \rhocv_{ik} \Lambda_{kj}^\rmv(\rho) - \sum_k \rhocv_{ik} \kappa_{kj}^\rmv,
\end{equation*}
and Equation (\ref{eq:Coulomb_cv}c) to 
\begin{eqnarray*}
\mean{[\vc{j} \ca{i},H^{\rmc-\rmv}]} 
& \WT & \sum_k \zeta_{ik}^\rmv(\rho) \rhocv_{kj} 
+ \sum_k \gamma_{ik}^{\rmc-\rmv}(\rho) \rhov_{kj} \\
&& - \sum_k \rhoc_{ik} \gamma_{kj}^{\rmc-\rmv}(\rho)
- \sum_k \rhocv_{ik} \zeta_{kj}^\rmc(\rho)
+ \sum_k \eta_{ik}^{\rmc-\rmv}\rhocv_{kj}.
\end{eqnarray*}

\subsection{Matrix formulation}
\label{sec:matrix}

We would like to cast the former results as
\begin{equation*}
\mean{\left[\begin{pmatrix}\cc{j}\ca{i} & \vc{j}\ca{i} \\
\cc{j}\va{i} & \vc{j}\va{i} \end{pmatrix},H^\rmC\right]} \WT [V^\rmC(\rho),\rho],
\end{equation*}
where 
\begin{equation*}
V^\rmC(\rho) = \begin{pmatrix}V^\rmc(\rho) & V^{\rmc-\rmv}(\rho) \\
V^{\rmv-\rmc}(\rho) & V^\rmv(\rho) \end{pmatrix},
\end{equation*}
which implies
\begin{eqnarray*}
\mean{[\cc{j}\ca{i},H^\rmC]} 
& \WT & V^\rmc(\rho) \rhoc + V^{\rmc-\rmv}(\rho) \rhovc - \rhoc V^\rmc(\rho) - \rhocv V^{\rmv-\rmc}(\rho), \\
\mean{[\vc{j}\va{i},H^\rmC]} 
& \WT & V^{\rmv-\rmc}(\rho) \rhocv + V^\rmv(\rho) \rhov - \rhovc V^{\rmc-\rmv}(\rho) - \rhov V^\rmv(\rho), \\
\mean{[\vc{j}\ca{i},H^\rmC]} 
& \WT & V^\rmc(\rho) \rhocv + V^{\rmc-\rmv}(\rho) \rhov - \rhoc V^{\rmc-\rmv}(\rho) - \rhocv V^\rmv(\rho).
\end{eqnarray*}
Identifying the coefficients computed in Sections \ref{sec:c}, \ref{sec:v}, and \ref{sec:cv}, we obtain that the result can indeed be cast as $[V^\rmC(\rho),\rho]$ where
\begin{eqnarray*}
V^\rmc(\rho) & = & \Lambda^\rmc(\rho) + \zeta^\rmv(\rho) + \eta^{\rmc-\rmv}, \\
V^{\rmc-\rmv}(\rho) & = & \gamma^{\rmc-\rmv}(\rho), \\
V^{\rmv-\rmc}(\rho) & = & \gamma^{\rmc-\rmv*}(\rho) = V^{\rmc-\rmv*}(\rho), \\
V^\rmv(\rho) & = & \Lambda^\rmv(\rho) + \zeta^\rmc(\rho) + \kappa^\rmv,
\end{eqnarray*}
and the various matrices have been defined by Equations \eqref{eq:Lambdac} to \eqref{eq:zetac}.

The evolution equation for the density matrix including also the free electron Hamiltonians, the interaction with a laser field, and Coulomb interaction can therefore be cast in a Liouville form
\begin{equation}
\label{eq:Bloch}
\rmi\hbar\partial_t \rho = [V(t,\rho(t)),\rho],
\end{equation}
where $V(t,\rho(t))=V_0(t)+V^\rmC(\rho(t))$ and $V_0(t)$ has been introduced in Equation \eqref{eq:basicBloch}.

\subsection{Energy shifts}

Recall (see Equation \eqref{eq:basicBloch}) the basic Bloch equation 
\begin{equation*}
\rmi\hbar\partial_t \rho = [V_0(t),\rho],
\end{equation*}
where
\begin{equation*}
V_0(t) = \left(\begin{array}{cc}
E_0^\rmc + \Re\bfE(t)\cdot\bfM^\rmc & \bfE(t)\cdot\bfM^{\rmc\rmv} \\
\bfE^*(t)\cdot\bfM^{\rmc\rmv*} & E_0^\rmv + \Re\bfE(t)\cdot\bfM^\rmv \\
\end{array}\right).
\end{equation*}
The symmetries in the definition of $\bfM^\rmc$ and $\bfM^\rmv$ imply that their diagonal entries are zero. 
On the contrary for $E_0^\rmc$ and $E_0^\rmv$ there are only diagonal entries.
In the evolution equation of $\rhoc_{ij}$, the $\rhoc_{ij}$ term only involves $E_0^\rmc$, and the other entries of $\rho$ play a role through $\bfM^\rmc$ and $\bfM^{\rmc\rmv}$.

We can therefore easily analyze the Coulomb contributions in terms of shifts on the free electron energies and off-diagonal terms.
Hence we compute
\begin{eqnarray*}
\delta\epsilon_i^\rmc(\rho) 
& = & 2 \sum_{(\a,\ap)\in(\Ic)^2} 
\left(\Rc_{i\a i\ap}-\Rc_{i\a\ap i}\right) \rhoc_{\ap\a} 
+ \sum_{(\a,\ap)\in(\Iv)^2} \Rcv_{i\a i\ap}\rhov_{\ap\a}
- \sum_{\beta\in\Iv} \Rcv_{i\beta i\beta}, \\
\delta\epsilon_i^\rmv(\rho) 
& = & 2 \sum_{(\a,\ap)\in(\Iv)^2} 
\left(\Rv_{i\a i\ap}-\Rv_{i\a\ap i}\right) \rhov_{\ap\a}
+ \sum_{(\a,\ap)\in(\Ic)^2} \Rcv_{\a i\ap i}\rhoc_{\ap\a} 
+ 2 \sum_{\beta\in\Iv} (\Rv_{\beta ii\beta}-\Rv_{\beta i\beta i}).
\end{eqnarray*}
We can define the energy shift matrices as $\delta E^\rmc(\rho)=\diag(\{\delta\epsilon_i^\rmc(\rho)\}_{i\in\Ic})$ and $\delta E^\rmv(\rho)=\diag(\{\delta\epsilon_i^\rmv(\rho)\}_{i\in\Iv})$. 
Hence $V(t,\rho(t))=E(\rho(t))+R(t,\rho(t))$, where $E(t,\rho(t))=\diag(E^\rmc(t,\rho(t)),E^\rmv(t,\rho(t)))$ and
\begin{equation*}
E^\rmc(\rho(t)) = E_0^\rmc + \delta E^\rmc(\rho(t))
\text{ and }
E^\rmv(\rho(t)) = E_0^\rmv + \delta E^\rmv(\rho(t)).
\end{equation*}

\subsection{Vanishing intra-band coherence assumption}

The model for quantum dots given in \cite{Gehrig-Hess02} has clearly been derived mimicking quantum well models as in \cite{Kuhn-Rossi92}. 
In quantum wells, electrons and holes can interact only if they "see each other" long enough, i.e. if they have (and are indexed by) the same wave vector. 
This leads morally to weakly coupled two-level systems.
Therefore, in \cite{Gehrig-Hess02}, the variables are the level populations and the inter-band coherences.
In our model, this means that intra-band coherences $\rhoc_{ij}$ and $\rhov_{ij}$ for $i\neq j$ are not considered.

We can therefore wonder what becomes of our model if we set intra-band coherences to zero.
First we notice that in the evolution equation of, e.g., $\rhoc_{ij}$, there is, e.g., a contribution of $V^\rmc_{ij} (\rhoc_{jj}-\rhoc_{ii})$.
Hence, even if intra-band coherences are initially zero, they are not always zero through the evolution with Equation \eqref{eq:Bloch}. 
Setting artificially intra-band coherences to zero therefore destroys the Liouville structure of the system. 

In our model, this hypothesis also changes the definition of $\Lambda^\rmc(\rho)$, $\zeta^\rmv(\rho)$ and $\Lambda^\rmv(\rho)$, which become
\begin{eqnarray*}
\Lambda_{ik}^\rmc(\rho) & = & 2 \sum_{\a\in\Ic} 
\left(\Rc_{i\a k\a}-\Rc_{i\a\a k}\right) \rhoc_{\a\a}, \\
\zeta_{ik}^\rmv(\rho) & = & \sum_{\a\in\Iv} \Rcv_{i\a k\a}\rhov_{\a\a}, \\
\Lambda_{ik}^\rmv(\rho) & = & 2 \sum_{\a\in\Iv} 
\left(\Rv_{i\a k\a}-\Rv_{i\a\a k}\right) \rhov_{\a\a}.
\end{eqnarray*}

\section{Mathematical analysis}
\label{sec:maths}

\subsection{Estimates on the density matrix}
\label{sec:estimates}

In the same way as in the usual Bloch case \cite{Bidegaray-Bourgeade-Reignier01}, the Liouville structure allows to state that the density matrix remains Hermitian through time evolution.
Its trace is conserved and it remains a positive semi-definite matrix.
Hence for a given electric field $\bfE(t)$, the elements of the density matrix are bounded, more precisely populations are bounded
\begin{equation*}
|\rho_{ii}^\rmc(t)|, |\rho_{jj}^\rmv(t)| \leq \Tr(\rho(t)) = \Tr(\rho(0)),
\hspace{1cm} \text{ for all } i\in\Ic \text{ and } j\in\Iv,
\end{equation*}
as well as coherences
\begin{equation*}
|\rho_{ij}^\rmc(t)| \leq \sqrt{\rho_{ii}^\rmc(t)\rho_{jj}^\rmc(t)}\leq \frac{\Tr(\rho(0))}2,
\hspace{1cm} \text{ for all } i,\ j\in\Ic,
\end{equation*}
and likewise
\begin{equation*}
|\rho_{jk}^\rmv(t)|, |\rho_{ij}^{\rmc-\rmv}(t)| \leq \frac{\Tr(\rho(0))}2,
\hspace{1cm} \text{ for all } i\in\Ic \text{ and } j,\ k\in\Iv.
\end{equation*}

\subsection{Coupling with the Maxwell equations}

The density matrix governed by Bloch equations in Section \ref{sec:Heisenberg} was only depending on time. 
We can now consider a collection of quantum dots which are scattered in space and interacting, not directly but through the interaction with an electromagnetic wave that propagates through the medium.
This can be modeled by a density matrix, that now depends on time and space, which is coupled with Maxwell equations for the laser field through the expression of polarization.
We therefore address the system
\begin{equation}
\label{eq:MB}
\left\{
\begin{aligned}
\mu\partial_t \bfH &= - \curl \bfE, \\
\eps\partial_t \bfE &= \curl \bfH - \partial_t \bfP, \\
\bfP & = \calN_\rmb \Tr(\bfM\rho), \\
\partial_t \rho & = - \frac\rmi\hbar [V(\rho),\rho],
\end{aligned}
\right.
\end{equation}
where all the variables depend on time and space in 3 dimensions: the electric and magnetic fields $\bfE$ and $\bfH$, the polarization $\bfP$, and the density matrix $\rho$. 
In Equation \eqref{eq:MB}, $\eps$ and $\mu$ denote the electromagnetic permittivity and permeability of the underlying medium. They both can depend on the space variable. The density of quantum boxes is given by $\calN_\rmb$.
The Bloch and Maxwell equations are coupled \textit{via} the polarization that involve the dipolar moment matrix $\bfM$. 

Such models have already been written and studied mathematically and numerically in a few physical contexts.
Here the specificity is the fact that the Bloch equation is nonlinear in $\rho$. 
Note that even in the case when $V$ does not depend on $\rho$, the full coupled model is already nonlinear since $V$ is affine in $\bfE$.

This system can be cast in the abstract setting of \cite{Dumas-Sueur12}.
In this paper, they introduce a general abstract setting able to treat both Maxwell--Landau--Lifschitz and classical Maxwell--Bloch equations.
In this setting, the electromagnetic field is supposed to exist in all space $\bbR^3$. 
Matter described by the density matrix is only occupying a subdomain $\Omega$ of $\bbR^3$.
The variables are gathered in one variable $U=(u,v)$, where $u=(u_1,u_2)=(\bfH,\bfE)$ and $v=\rho$. 
The variable $u$ can be viewed as 6 real variables and variable $v$ as $d^2$ real variables.
This variable $U$ is supposed to be in $\bfL^2=L^2(\bbR^3;\bbR^6)\times L^2(\Omega;\bbR^{d^2})$. 
The abstract system reads
\begin{equation}
\label{eq:abstract}
\left\{
\begin{aligned}
(\partial_t+B)u &= (\kappa^{-1}\cdot l) F(\bar v, u), 
&& \text{ for }x\in\bbR^3, \\
\partial_t v &= F(v,u), 
&& \text{ for }x\in\Omega.
\end{aligned}
\right.
\end{equation} 
In this formulation $\kappa(x)=(\kappa_1(x),\kappa_2(x))=(\mu,\eps)$.
It is uniformly positive as needed in \cite{Dumas-Sueur12}.
Let $H_{\rm curl}$ be the space of functions $f\in L^2(\bbR^3;\bbR^3)$ with $\curl f\in L^2(\bbR^3;\bbR^3)$.
The linear differential operator $B$ is defined on $H_{\rm curl}\times H_{\rm curl}$ by $B(u_1,u_2)=(\kappa_1^{-1}\curl u_2,-\kappa_2^{-1}\curl u_1)$. 
The variable $v$ is extended by $\bar v$ on the whole $\bbR^3$ and is zero outside $\Omega$. We can identify $l_1=0$, $l_2=-\calN_\rmb\Tr(\bfM\cdot)$, and $F(v, u)=- \frac\rmi\hbar [V(\rho),\rho]$.
The system \eqref{eq:MB} verifies the hypotheses given in \cite{Dumas-Sueur12}, namely,
\begin{itemize}
\item $F$ is affine in $u$: $F(v,u)=F_0(v)+F_1(v)u$,
\item for $j=0,1$, $F_j(0)=0$,
\item for all $R>0$ there exists $C_F(R)$ such that for all $v\in B_R$ (ball of radius $R$ in $\bbR^d$), $|F_j(v)|+|\partial_vF_j(v)|\leq C_F(R)$,
\item there exists $K\geq0$ such that for all $(u,v)\in\bbR^6\times \bbR^{d^2}$, $F(v,u)\cdot v\leq K|u|^2$.
\end{itemize}
We have in particular used the $L^\infty$ bounds of Section \ref{sec:estimates}, more precisely, we look for 
\begin{equation}
\label{eq:finiteenergy}
v\in L^\infty((0,\infty);L^\infty(\Omega;\bbR^{d^2})).
\end{equation}
Besides we suppose to have at time $t=0$ the conditions
\begin{equation}
\label{eq:div}
\div(\kappa_ju_j-l_j\bar v)= 0, \hspace{1cm} \text{ for } j=1,2,
\end{equation}
which are indeed the physical conditions $\div (\mu\bfH)=0$ and $\div(\eps \bfE+\bfP)=0$ \cite{Dumas05}.
The structure of Equation \eqref{eq:abstract} ensures that this condition holds for all time if it is valid at the initial time.

\subsection{Cauchy problem}

In this section, we state without proof the results obtained in \cite{Dumas-Sueur12} and that we can apply to our context. 
The first result addresses the existence of global finite energy solutions.

\begin{Theorem}[Theorem 3, \cite{Dumas-Sueur12}]
\label{th:existence}
For any initial data $U_0=(u_0,v_0)\in L^2(\bbR^3;\bbR^6)\times(L^2(\Omega;\bbR^{d^2})\cap L^\infty(\Omega;\bbR^{d^2}))$ satisfying $\eqref{eq:div}$, there exists $U\in\calC([0,\infty);\bfL^2)$ which is a solution to \eqref{eq:abstract}--\eqref{eq:div}, and satisfies the finite energy condition \eqref{eq:finiteenergy}. 
Moreover, for all $T>0$, there exists a constant $C$ that only depends on $T$, $F$, $l$ and $\|v_0\|_{L^\infty}$, such that for all $t\in[0,T]$, $\|U(t)\|_{\bfL^2} \leq C \|U_0\|_{\bfL^2}$.
\end{Theorem}

To have a uniqueness result, we need some regularity on $\eps$ and $\mu$. 
Usually in physical contexts, $\eps$ and $\mu$ may be discontinuous across the boundary of $\Omega$, but we do not know how to tackle with this problem.
We hence assume that
\begin{equation}
\label{eq:smooth}
\Omega \text{ is bounded and } \kappa_i-1 \in \calC^\infty_{\bar\Omega}(\bbR^3), \text{ for } i=1,2,
\end{equation} 
the space of $\calC^\infty$ functions on $\bbR^3$ with compact support included in $\bar\Omega$, which means in particular that $\kappa_i$ is 1 outside $\Omega$ and the transition across the boundary is smooth.
We also assume that the initial data for the electromagnetic wave is smooth enough.
 
\begin{Theorem}[Theorem 5, \cite{Dumas-Sueur12}]
Under the assumptions of Theorem \ref{th:existence} and \eqref{eq:smooth}, and assuming that $\curl u_{0i}\in L^2(\bbR^3)$ for $i=1,2$, there exists only one solution to \eqref{eq:abstract}--\eqref{eq:div} with initial data $U_0$, given by Theorem \ref{th:existence}. It satisfies $\curl u_i\in\calC([0,\infty);L^2(\bbR^3))$ for $i=1,2$.
\end{Theorem}
This stems from the fact that $l_1=0$ and $F$ does not depend on $u_1$ but only $u_2$.

\section{Numerical experiments}
\label{sec:numerics}

\subsection{Self-Induced Transparency}

Self-Induced Transparency (SIT) is a typical two-level phenomenon: using a light pulse which is resonant with the transition, absorption and stimulated emission are combined to obtain exact population inversion and an unchanged electric field. This phenomenon has been predicted theoretically and confirmed experimentally \cite{Allen-Eberly87,McCall-Hahn67,Gibbs-Slusher70}. 

The propagating field is a pulse given by
\begin{equation*}
E(t,z) = \calE(t,z) \sin(\omega_0(t-z/v)), 
\end{equation*}
where $v$ is the velocity of the pulse, $\omega_0$ is both the center frequency of the pulse and the transition frequency of the medium, and $\calE(t,z)$ is the pulse envelope. 
It is shown that the envelope is not reshaped by the medium, only if it is a symmetric hyperbolic secant
\begin{equation*}
\calE(t,z) = \calE_0 \sech\left(\frac{t-z/v}\tau\right), \text{ where }\calE_0=\frac2{m\tau}.
\end{equation*}
In this expression, $\tau$ is the pulse duration and $m=M/\hbar$, where $M$ is the dipolar moment associated to the transition.
According to the Area Theorem \cite{McCall-Hahn67}, the medium undergoes $k$ exact inversions if  
\begin{equation*}
A = m \int_{-\infty}^\infty \calE(t,z) dt = k\pi.
\end{equation*}
The corresponding pulse is called a $k\pi$-pulse.
It is easy to compute that 
\begin{equation*}
A = m\tau\calE_0 \left[\arctan(\sinh t)\right]_{-\infty}^\infty = m\tau\calE_0 \pi,
\end{equation*}
and hence a $k\pi$-pulse is obtained for the amplitude $\calE_0=k\hbar/M\tau$.
For our test-cases, we will use a $2\pi$-pulse, for which the medium is inverted and goes eventually to its original state. 
The return to the initial state is an easy-to-check criterion to validate numerical approaches, as has been already done in \cite{BidegarayFesquet06a,Ziolkowski-Arnold-Gogny95}.

\subsection{Adaption to the quantum dot context}

In this paper we want in particular to investigate the validity of the vanishing intra-band coherence assumption. 
To this aim we need a minimum of three levels and we therefore adapt the SIT experiment to our framework.
We absolutely do not claim that SIT has any practical application in the quantum dot context, but only choose this test-case because of the easiness to interpret the results.

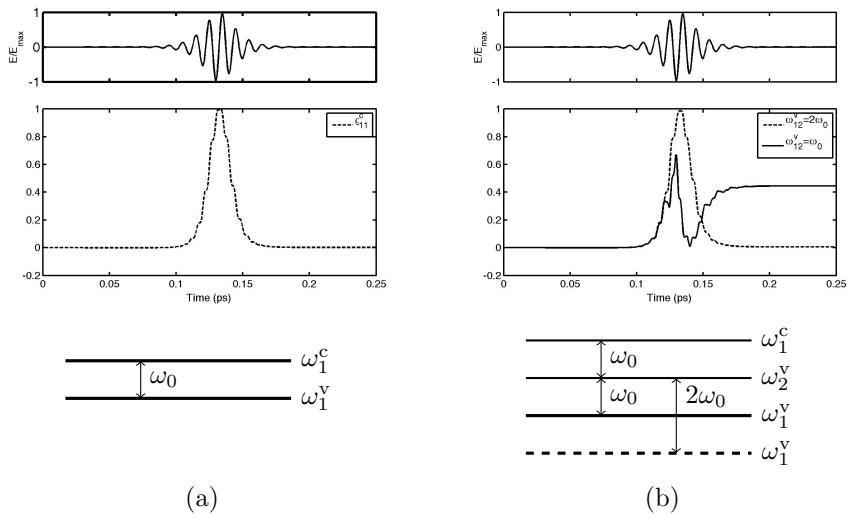
\begin{figure}[H]
\begin{center}
\begin{tabular}{ccc}
\includegraphics[width=.47\textwidth]{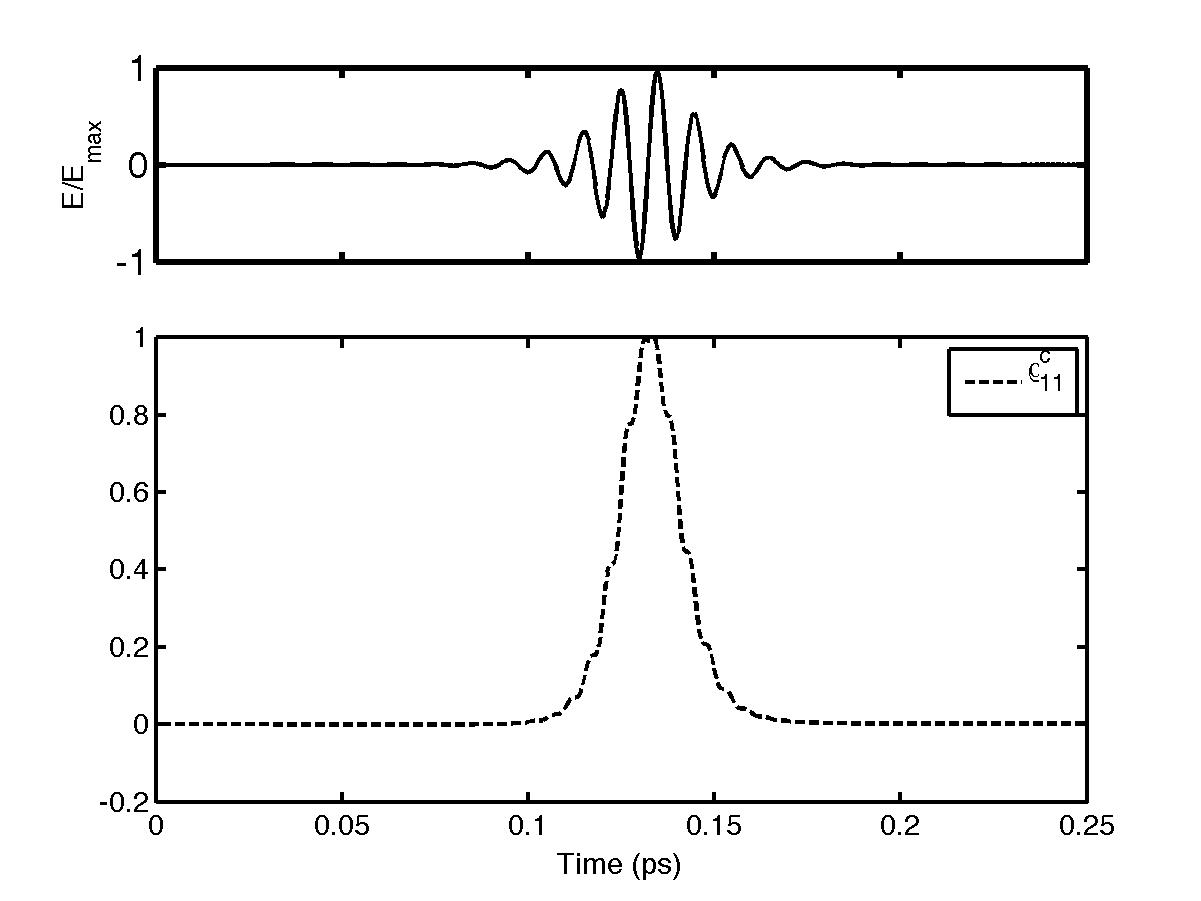} 
& \includegraphics[width=.47\textwidth]{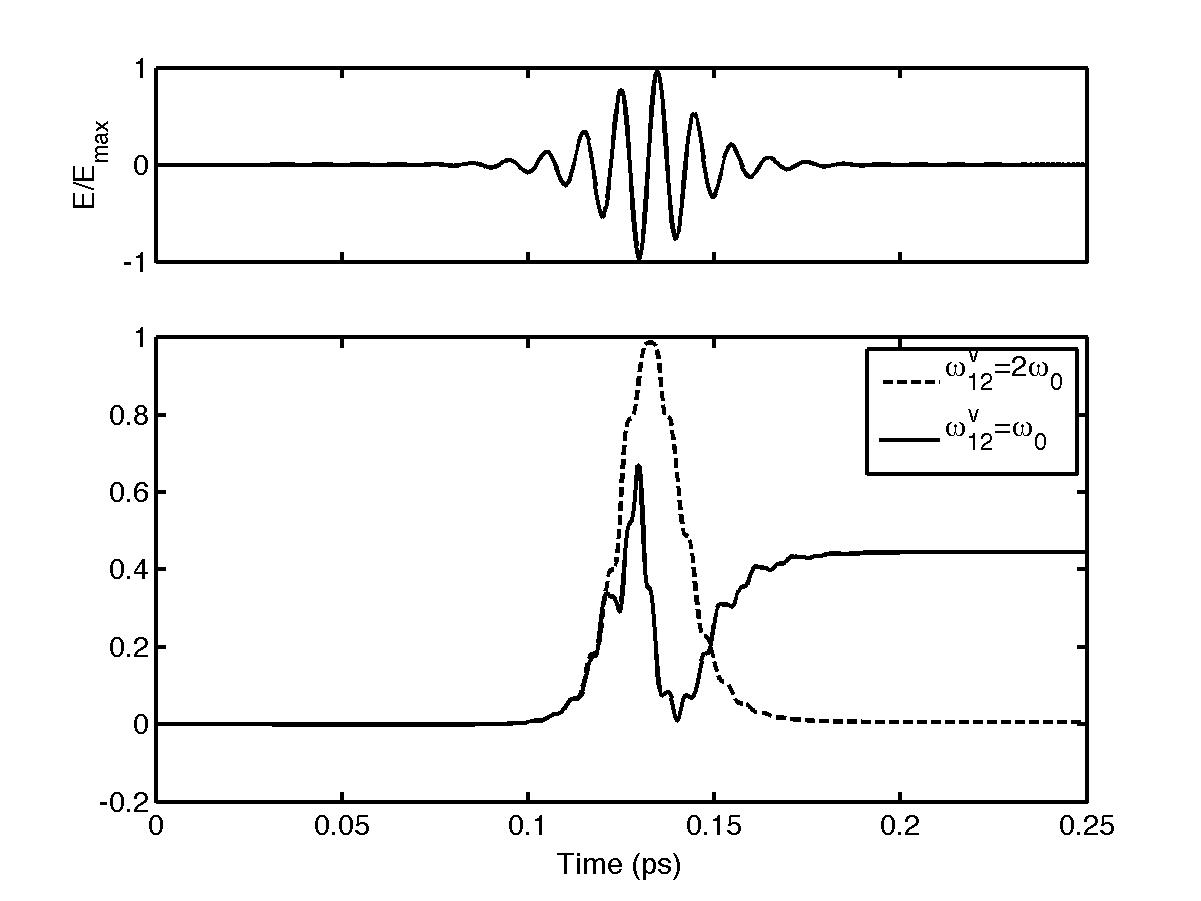} \\
\begin{tikzpicture}
\draw[very thick](0,.5) -- (3,.5) node[right] {$\omega_1^\rmc$};
\draw[very thick](0,0) -- (3,0) node[right] {$\omega_1^\rmv$};
\draw[<->](1,0) -- node[right] {$\omega_0$} (1,.5);
\draw[color=white](0,-1) -- (3,-1);
\end{tikzpicture}
&
\begin{tikzpicture}
\draw[thick](0,.5) -- (3,.5) node[right] {$\omega_1^\rmc$};
\draw[thick](0,0) -- (3,0) node[right] {$\omega_2^\rmv$};
\draw[<->](1,0) -- node[right] {$\omega_0$} (1,.5);
\draw[very thick](0,-.5) -- (3,-.5) node[right] {$\omega_1^\rmv$};
\draw[<->](1,-.5) -- node[right] {$\omega_0$} (1,0);
\draw[dashed,very thick](0,-1) -- (3,-1) node[right] {$\omega_1^\rmv$};
\draw[->](2,-.5) -- node[right] {$2\omega_0$} (2,0);
\draw[->](2,-.5) -- (2,-1);
\end{tikzpicture}
 \\
(a) & (b) \\
\end{tabular}
\end{center}
\caption{\label{fig:SIT}Adaption of the SIT test case to the quantum dot context. (a) Original two-level case; (b) 2 three-level test cases.}
\end{figure}

In Figure \ref{fig:SIT}(a), the original two-level test case is represented, for which there is a single conduction level and a single valence level, separated by the energy corresponding to the field frequency.
The upper plot represents the (normalized) time-evolution of the electric field. 
The time-evolution of the population of the initially empty conduction level is given by the lower curve.
We observe that the medium undergoes two complete population inversions.

In Figure \ref{fig:SIT}(b), we have two three-level test cases with a single conduction level and two valence levels.
In the first place we do not take Coulomb interaction into account.
In the first test case (represented by solid lines both on the plot and on the scheme) the transition between the two valence levels is also resonant with the field and this destroys the SIT phenomenon. 
It suffices to get both valence levels far apart enough (e.g. $2\omega_0$ as in the second case, represented by dashed lines) to recover SIT. 
We use this last configuration as basis test case for the following numerical experiments.

\subsection{Numerical features}

The simulations are performed using a code based on a finite difference Yee scheme and a relevant choice for the time discretization of the Bloch equation (see \cite{Bidegaray03}). 
It allows to keep the good properties of the original Yee scheme: second order and explicitness. 
A splitting scheme, first described in \cite{Bidegaray-Bourgeade-Reignier01}, allows to preserve positiveness at the discrete level.
It is strongly based on the exact solution given by Equation \eqref{eq:exact}. 
It has been adapted to include also Coulomb interaction, still preserving positiveness, but at the cost of a loss of approximation order, which becomes one. 

Integrating the zero intra-band coherences assumption is \textit{a priori} a problem since it destroys the Liouville structure and an exact solution is no more available. 
It is however possible to solve a Liouville-like equation and set artificially intra-band coherences to zero.
This adds a step at each time iteration but allows to preserve the general structure of the numerical code.

To determine the right envelope amplitude for numerics, we use the argument of \cite{Ziolkowski-Arnold-Gogny95}: in practice the input pulse is cut off on an interval $t\in[-10\tau,10\tau]$, therefore the numerical area is  
\begin{equation*}
A_\rmn = m\tau\calE_0 \left[\arctan(\sinh t)\right]_{-10}^{10} = m\tau\calE_0 (0.999942\pi),
\end{equation*}
which slightly changes the value of $\calE_0$.

\subsection{Results}

\subsubsection{Impact of Coulomb terms}

To include Coulomb interaction in the SIT test case, we have to give values to the coefficients given by Equation \eqref{eq:R}. 
Their exact computation is not in the scope of the present paper.
We choose to take them of the same order $R_0$, taking into account the symmetries described by Property \ref{prop:R}, but not equal (which would lead to vanishing $\Lambda^\rmc$, $\Lambda^\rmv$ and $\kappa^\rmv$). The test is performed using the full Coulomb terms (no vanishing intra-band coherence assumption).

\begin{figure}[t]
\begin{center}
\begin{tabular}{cc}
\includegraphics[width=.47\textwidth]{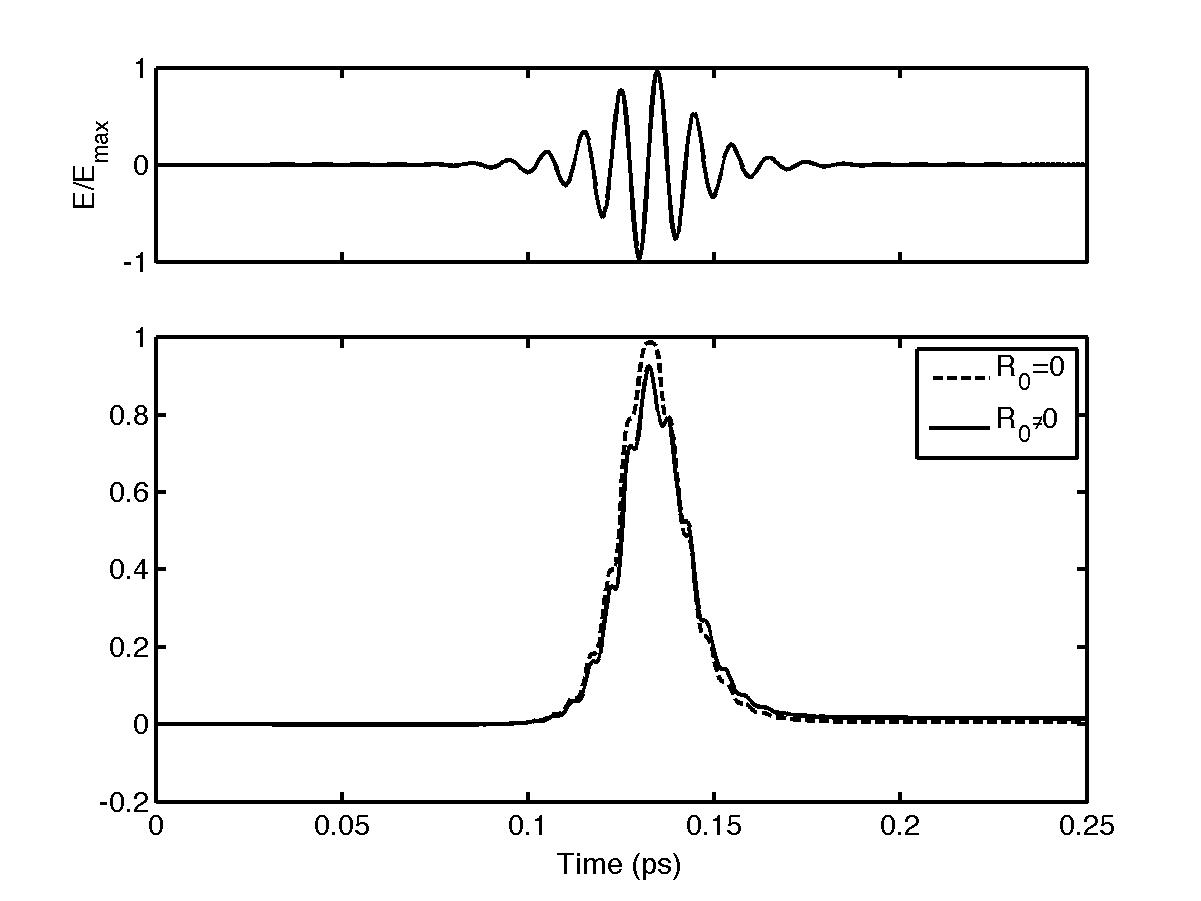} 
& \includegraphics[width=.47\textwidth]{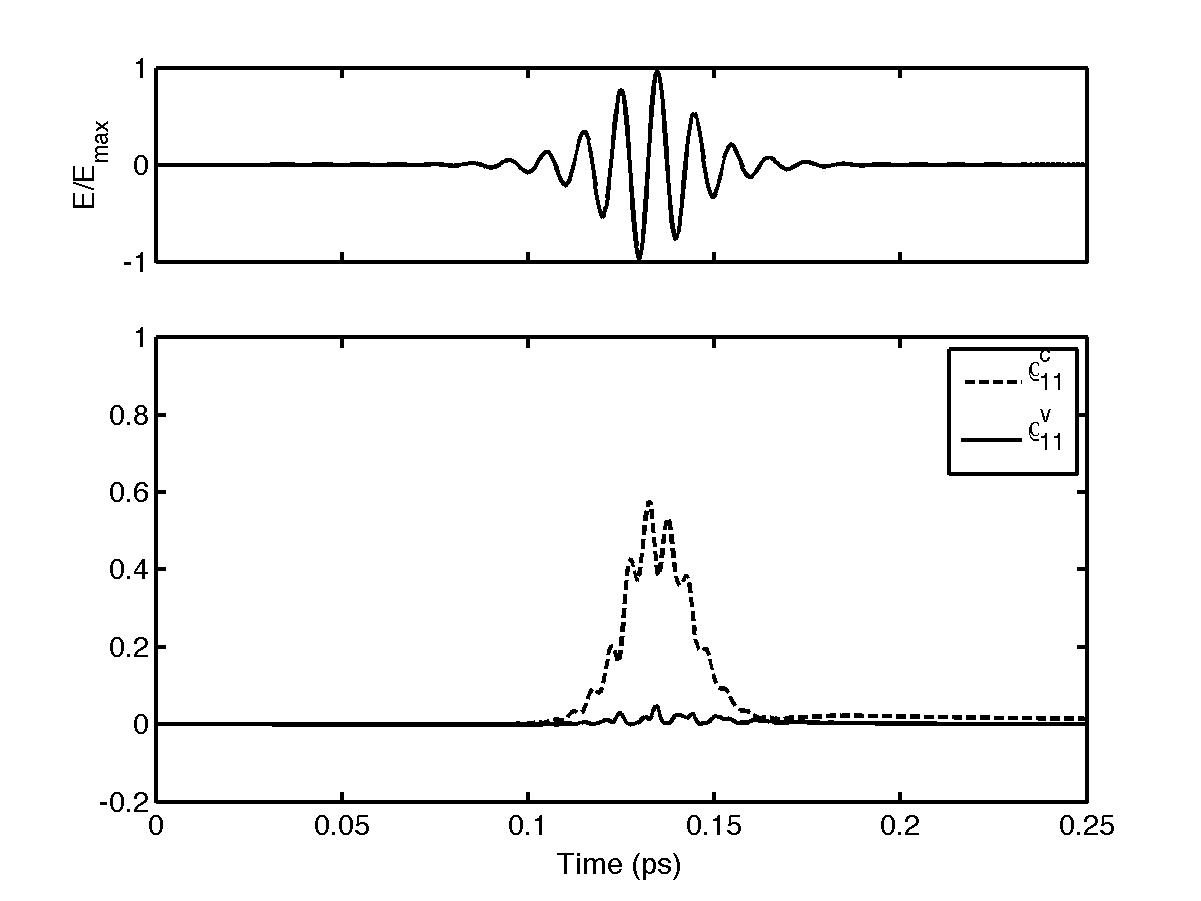} \\
(a) $R_0 = 10^{-21}$ & (b) $R_0 = 3\times10^{-21}$ 
\end{tabular}
\end{center}
\caption{\label{fig:Coulomb}Impact of Coulomb terms for two interaction strengths. (a) $R_0 = 10^{-21}$ -- comparison with the Coulomb-free case; (b) $R_0 = 3\times10^{-21}$ -- excitation of the conduction level and of the first valence level.}
\end{figure}

For small values of $R_0$ (see the evolution of $\rho_{11}^\rmc$  described for $R_0=10^{-21}$ in Figure \ref{fig:Coulomb}(a), solid plot) SIT is only slightly affected. 
In this figure, the dashed plot corresponds to the reference case ($R_0=0$) and is the same as the dashed plot of Figure \ref{fig:SIT}(b).
The effect is clearer for stronger values of $R_0$ (e.g.,\break $R_0 = 3\times 10^{-21}$ in Figure \ref{fig:Coulomb}(b)). 
The total inversion is prevented by Coulomb interaction.
Although inversion is not complete, the return to zero of $\rho_{11}^\rmc$  is observed in this test case.
We notice that the first valence level (which is not supposed to take part in the SIT experiment) is slightly populated, which makes this test case not really a two-level experiment.

\subsubsection{Impact of vanishing intra-band coherences}

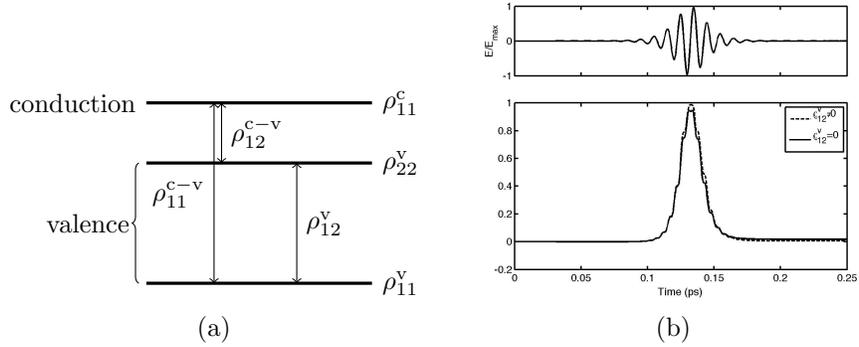
\begin{figure}[H]
\begin{center}
\begin{tabular}{cc}
\begin{tikzpicture}
\usetikzlibrary{snakes}
\draw[very thick](0,.8) node[left] {conduction} -- (3,.8) node[right] {$\rho_{11}^\rmc$};
\draw[very thick](0,0) -- (3,0) node[right] {$\rho_{22}^\rmv$};
\draw[<->](1,0) -- node[right] {$\rho_{12}^{\rmc-\rmv}$} (1,.8);
\draw[very thick](0,-1.6) -- (3,-1.6) node[right] {$\rho_{11}^\rmv$};
\draw[<->](2,0) -- node[right] {$\rho_{12}^\rmv$} (2,-1.6);
\draw[<->](.9,-1.6) -- node[left] {$\rho_{11}^{\rmc-\rmv}$} (.9,.8);
\draw[snake=brace] (-.1,-1.6) -- node[left] {valence} (-.1,0);
\end{tikzpicture}
& \includegraphics[width=.47\textwidth]{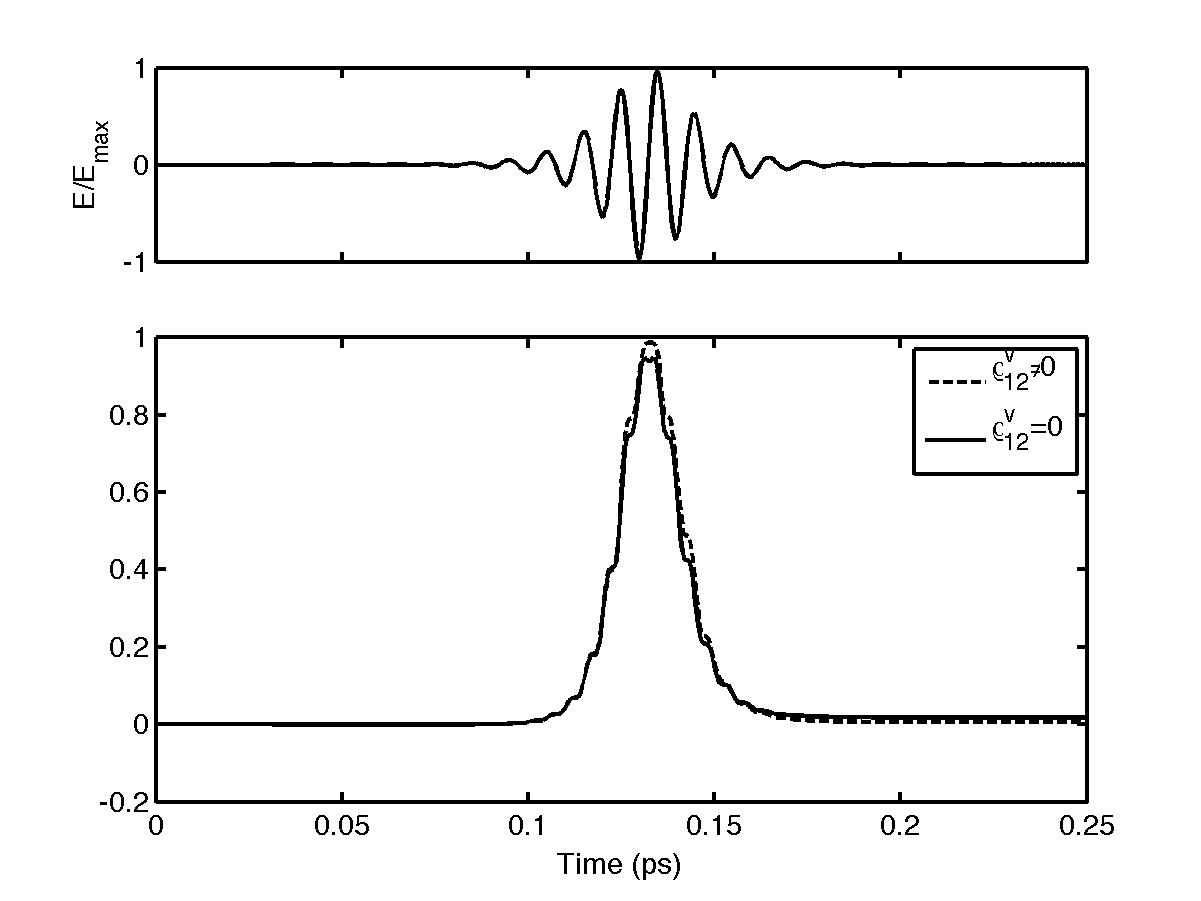} \\
(a) & (b) 
\end{tabular}
\end{center}
\caption{\label{fig:vanish1}Impact of vanishing intra-band coherences on the Coulomb-free model.}
\end{figure}

We first test the impact of the vanishing intra-band coherence assumption on the Coulomb-free model. 
We always use the same experimental setting (see Figure \ref{fig:vanish1}(a)) and this assumption amounts to taking $\rho_{12}^\rmv=0$.

The result is displayed in Figure \ref{fig:vanish1}(b).
The final equilibrium state for matter is slightly changed and $\rho_{11}^c$, which is given by the solid curve, does not eventually return to zero. Inversion is not total.

Now we combine both Coulomb interaction and the vanishing intra-band coherence assumption. If $R_0$ is low (e.g., $10^{-21}$), the result is not much affected by this assumption and is essentially the same as that plotted on Figure \ref{fig:Coulomb}(a).

\begin{figure}[t]
\begin{center}
\includegraphics[width=.9\textwidth]{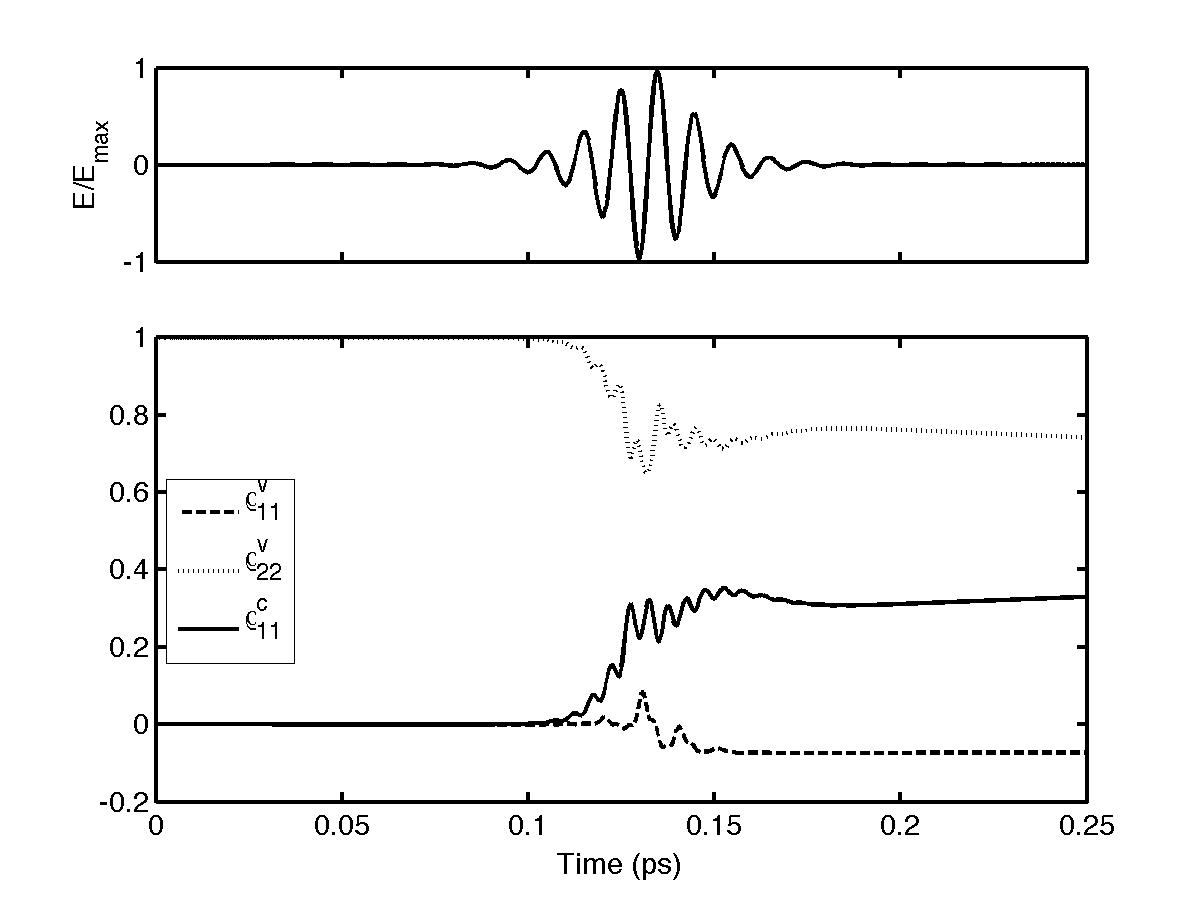} 
\end{center}
\caption{\label{fig:vanish2}Impact of vanishing intra-band coherences on the full model for $R_0 = 3\times10^{-21}$.}
\end{figure}

In the case when $R_0 = 3\times10^{-21}$, we see in Figure \ref{fig:vanish2} that the final equilibrium state is not physical ($\rho_{11}^\rmv<0$). 
This is due to the destruction of the Liouville structure. 
We can easily explain why this is not observed when $R_0$ is low.
When we are close to the SIT experiment we have a typical two-level phenomenon: $\rho_{11}^\rmv\simeq0$ during the whole experiment.
For a two-level system, the positiveness of the density matrix is equivalent to 
\begin{itemize}
\item the positiveness of each diagonal term (populations),
\item the estimation of coherence by populations, here: $|\rho_{12}^{\rmc-\rmv}|^2\leq\rho_{11}^\rmc\rho_{22}^\rmv$, if the second valence level would be the only relevant one (see \cite{Bidegaray-Bourgeade-Reignier01}).
\end{itemize}
Setting intra-band coherences to zero within the numerical process does not affect these properties, and the iteration used in the proof of the positiveness of the density matrix applies.

But for a three-level system (and the case $R_0 = 3\times10^{-21}$ is a true three-level case), the positiveness of the matrix involves some more properties, which are affected by setting $\rho_{12}^\rmv$ to zero.
Although trace is still conserved, the positiveness of the population is affected.
We would of course have the same result with a dedicated code where intra-band coherences would simply not be computed.
Besides the effect is clear enough in Figure \ref{fig:vanish2} not to be attributed to simple round-off errors (there are only 600 time-steps in this computation).
The conclusion is that even if intra-band coherences seem not to be very relevant for some physical applications, it is very important to include them in the mathematical description and in the numerical computation to keep the natural mathematical structure of the density matrix.

\begin{Remark}
In absence of electromagnetic field the evolution equation for the conduction electrons is reduced to 
\begin{equation*}
\rmi \hbar \partial_t \rho_{11}^\rmc = 0.
\end{equation*}
This is only due to the fact that there is only one conduction level in our test case, and does not depend on the intra-band coherence vanishing assumption or on specific values of the Coulomb coefficients.
Hence, when the population of the conduction level has been set into a non-physical state (and this is due to the intra-band coherence vanishing assumption), it remains in this state for ever.
\end{Remark}

\section{Conclusion}

In this paper, Bloch-type equations have been derived considering Coulomb effects in quantum dots.
We have shown analytically and numerically that Coulomb effects are not negligible in some quantum dot structures, and we have given the link between mathematical properties and physical relevancy of the Bloch model and more specifically in the treatment of intra-band coherences.
Then this model has been coupled with the description of laser propagation in the quantum dot structures, leading to a Maxwell-Bloch system for which we have studied the global Cauchy problem.
This system has been implemented numerically and simulations have been performed on a self-induced transparency test-case. 
In particular, we have tested the impact of Coulomb parameters and intra-band coherences. 
We have illustrated numerically that the modification of the equation structure when intra-band coherences are neglected can lead to non-physical solutions.
Further work will include additional effects in the same Bloch-type framework.

\subsection*{Acknowledgements}
LJK is partner of the LabEx PERSYVAL-Lab (ANR-11-LABX-0025-01) funded by the French program Investissement d’avenir. The authors wish to thank Eric Dumas for fruitful discussions.

\end{document}